\documentclass[12pt]{article}
\usepackage{graphicx}
\usepackage{epsfig}
\newcommand{\beq}{\begin{equation}}
\newcommand{\eeq}{\end{equation}}
\newcommand{\beqa}{\begin{eqnarray}}
\newcommand{\eeqa}{\end{eqnarray}}
\newcommand{\beqar}{\begin{eqnarray*}}
\newcommand{\eeqar}{\end{eqnarray*}}

\newcommand{\labell}[1]{\label{#1}} %\qquad_{#1}} %{\label{#1}}
\newcommand{\reef}[1]{(\ref{#1})}

\newcommand{\eg}{{\it e.g.,}\ }
\newcommand{\ie}{{\it i.e.,}\ }

\begin{document}

\setlength{\unitlength}{1mm}

\thispagestyle{empty}
\rightline{\small hep-th/0402149 \hfill}
\vspace*{3cm}

\begin{center}
{\bf \Large Rotating Circular Strings,}\\
{\bf \Large and}\\ 
{\bf \Large Infinite Non-Uniqueness of Black Rings}\\
\vspace*{1.7cm}

{\bf Roberto Emparan}\footnote{E-mail: {\tt emparan@ub.edu}}

\vspace*{0.2cm}

{\it Instituci\'o Catalana de Recerca i Estudis Avan\c cats (ICREA)}\\
{\it and}\\
{\it Departament de F{\'\i}sica Fonamental, and}\\ 
{\it C.E.R. en
Astrof\'{\i}sica, F\'{\i}sica de Part\'{\i}cules i Cosmologia,}\\
{\it Universitat de Barcelona, Diagonal 647, E-08028 Barcelona, Spain}\\[.5em]

\vspace{1.3cm} {\bf ABSTRACT}  
\end{center} 

\noindent We present new self-gravitating solutions in five dimensions
that describe circular strings, \ie rings, electrically coupled to a
two-form potential (as \eg fundamental strings do), or to a dual
magnetic one-form. The rings are prevented from collapsing by rotation,
and they create a field analogous to a dipole, with no net charge
measured at infinity. They can have a regular horizon, and we show that
this implies the existence of an infinite number of black rings, labeled
by a continuous parameter, with the same mass and angular momentum as
neutral black rings and black holes. We also discuss the solution for a
rotating loop of fundamental string. We show how more general rings
arise from intersections of branes with a regular horizon (even at
extremality), closely related to the configurations that yield
the four-dimensional black hole with four charges. We reproduce
the Bekenstein-Hawking entropy of a large extremal ring through a
microscopic calculation. Finally, we discuss
some qualitative ideas for a microscopic understanding of neutral
and dipole black rings.

\vfill \setcounter{page}{0} \setcounter{footnote}{0}
\newpage

\setcounter{equation}{0}
\section{Introduction}
\labell{intro}

Take a neutral black string in five dimensions, constructed as the direct
product of the Schwarzschild solution and a line, so the geometry of the
horizon is ${\bf R}\times S^2$. Imagine bending this string to form
a circle, so the topology is now $S^1\times S^2$. In principle this
circular string tends to contract, decreasing the radius of the $S^1$,
due to its tension and gravitational self-attraction, but we can make the
string rotate along the $S^1$ and balance these forces against the
centrifugal repulsion. Then we end up with a neutral rotating black
ring. Ref.~\cite{ER} obtained it as an explicit solution of
five-dimensional vacuum General Relativity.

This heuristic construction was first suggested surprisingly long ago in
\cite{MP}, but several of the most important features of the neutral
black ring could hardly have been anticipated without the explicit
solution. For fixed mass, the spin of a five-dimensional rotating black
hole of spherical topology is bounded above \cite{MP}, whereas the spin
of the black ring is bounded {\it below}. But the ranges of existence of
both sorts of objects overlap, and where they do, one actually finds {\it
two} black rings, in addition to the black hole, all with the same mass
and spin. This triplicity of solutions implies the non-uniqueness of
five-dimensional black holes, and can be regarded as a sort of hair,
even if it makes a rather thin wig.

A string can naturally couple to a two-form potential $B_{\mu\nu}$, and
be an electric source for it, a familiar example being the fundamental
string. Alternatively, its electric-magnetic dual in five dimensions
would be a string magnetically charged under a one-form potential
$A_\mu$. Take one such string and bend it into circular shape, balancing
it again by appropriately spinning the ring. This configuration will
have a non-trivial gauge field, but now the net charge will be zero.
Actually, as we shall ellaborate shortly, it is appropriate to view it
as a dipole, with zero net charge but with a non-vanishing local
distribution of charge. So we may refer to the rotating circular strings
as rotating dipole rings. These were conjectured to exist in
ref.~\cite{tubular,harv,EE}, and they are the subject of this paper.

The dipole field allows black rings to sport much thicker hair, and
indeed it realizes a more drastic {\it infinite} non-uniqueness: the
only conserved asymptotic charges of these rings are their mass and
spin, but they support a gauge field labeled by parameters within a
continuous range of values. This possibility was first anticipated in
\cite{harv}, and it will be realized below.

Since these solutions appear quite naturally in the supergravity
description of string/M-theory at low energies, they give us a new
perspective on the role of black rings within string theory, in a guise
different from the one studied in \cite{EE}. We begin to explore it in
this paper, and in particular provide the first precise microscopic
calculation of the entropy of a black ring.

\medskip

The structure of the paper is the following: Since the study of the
solutions is somewhat technical, before plunging into the details we
introduce, in the next section, the basic concepts and describe, in a
graphical manner, the main features of dipole rings. In section
\ref{dipolerings}, after a short description of the neutral rotating
ring, we give the explicit form of the dipole solutions and compute
their properties. In section \ref{floop} we address the issues raised by
the solution that describes a rotating loop of fundamental string, and
discuss qualitatively some features of dipole rings from a microscopic
string viewpoint. In section \ref{brane} we explain how black rings
appear in triple intersections of branes, with the rotation arising from
momentum running along the intersection. We also reproduce the
Bekenstein-Hawking entropy of the extremal black ring, to
next-to-leading order at large radius, via a statistical-mechanical
counting of string states. We conclude in section \ref{discuss} with a
discussion of the results and some qualitative considerations towards a
broader microscopic view of black rings. The appendix shows how the
Myers-Perry (MP) black hole \cite{MP} is recovered from the solutions in
section \ref{dipolerings}.

\setcounter{equation}{0}
\section{Setup and summary of properties}
\label{setup}

We construct dipole ring solutions for a number of theories, the
simplest of which are the five-dimensional Einstein-Maxwell-dilaton
theories
\beq\label{emdaction}
I=\frac{1}{16\pi G}\int d^5x\sqrt{-g}\left(
R-\frac{1}{2}(\partial\phi)^2-\frac{1}{4}e^{-\alpha\phi} F^2\right)\,.
\eeq
The conventional Einstein-Maxwell theory is recovered when the dilaton
decouples, \ie $\alpha=0$. One often considers the addition of a
Chern-Simons term to this theory, as required by minimal
supergravity in five dimensions. It turns out that the Chern-Simons term is of no
consequence to the solutions in this paper, and therefore the dipole
rings with $\alpha=0$ are also solutions of minimal five-dimensional
supergravity. Then the uniqueness theorem of supersymmetric black holes
in this theory \cite{harv} directly implies that none of the black rings
of the non-dilatonic theory can be a supersymmetric solution.

A ring is a circular string, and in the five-dimensional theories of
\reef{emdaction}, strings act as line sources of a magnetic field, \ie
they can be thought of as linear distributions of magnetic monopoles. So
the field of a magnetic black ring, outside the horizon, can be
described as being produced by a circular configuration of magnetic
monopoles. It is important to realize that, even if there is a local
distribution of charge, the total magnetic charge is zero \cite{dggh}.
The reason is that, in order to compute the magnetic charge in five
dimensions, one has to specify a two-sphere that encloses a point of the
string, {\it and} a vector tangent to the string. Since this vector
changes orientation halfway around the ring, the total charge on the
ring is zero. This can also be seen as the fact that on a 4D slice that
cuts the ring at diametrically opposite points, one finds opposite
magnetic charges. So the ring is analogous to a dipole.

We will also consider the electric dual of these
solutions. The transformation
\beq
\tilde\phi= -\phi,\qquad H= e^{-\alpha\phi}\ast F\,,
\eeq
where $H=dB$ is a three-form field strength, maps the theory
\reef{emdaction} to
\beq\label{e2daction}
I=\frac{1}{16\pi G}\int d^5x\sqrt{-g}\left(
R-\frac{1}{2}(\partial\tilde\phi)^2-\frac{1}{12}e^{-\alpha\tilde\phi} H^2\right)\,.
\eeq
%The two-form potential for a straight string extending along the
%direction $z$ is of electric type $B_{tz}$. For a dipole ring where
%$\psi$ is the angle along the circular direction of the ring the
%potential would be $B_{t\psi}$.

It will be convenient to express the dilaton coupling as
\beq\label{alphan}
\alpha^2=\frac{4}{N}-\frac{4}{3}\,,\qquad 0< N\leq 3\,,
\eeq
since the values $N=1,2,3$ are of particular relevance to
string and M-theory. In these cases the solution can be regarded as an
$N$-fold intersection of branes, which typically wrap an internal space
(see \eg \cite{tseyt}).
$N=3$ yields the non-dilatonic theory, but another case of
particular interest is $N=1$, since then the action \reef{e2daction}
can be interpreted as the NS sector of low energy string theory (in Einstein
frame), and it contains the fundamental string as a solution. The
dilaton $\sigma$ of string theory is in this case
$\sigma=\sqrt{\frac{3}{8}}\:\tilde\phi$ and the string metric
$g^{(s)}_{\mu\nu}=e^{\sqrt{\frac{2}{3}}\:\tilde\phi}g_{\mu\nu}$. Via
dualities the solution is related as well
to other single-brane configurations.

One can define a ``local charge" for the string solutions of
\reef{e2daction} as\footnote{Local, in the sense of corresponding to a
localized source of the gauge field, which may not give rise to a
net charge, but not referring to the local (as opposed to global) character
of the gauge
symmetry.}
\beq\label{elcharge}
{\cal Q}=\frac{1}{4\pi}\int_{S^2}e^{-\alpha\tilde\phi}\ast H\,,
\eeq
where the $S^2$ encloses a point along the string, and give it a sign
according to a choice of orientation along the string. To see the
meaning of this charge, consider first the case of a straight fundamental string.
This is infinitely long, but it can be made finite by requiring the
spatial direction parallel to the string to form a compact circle. The
string has a
topological winding number that is proportional to the local charge
\reef{elcharge}, 
\beq\label{winding}
n=\frac{\pi\alpha'}{2G}{\cal Q}\,,
%\left(\frac{2\pi}{G}\right)^{1/3}{\cal Q}\,,
\eeq
and this is a conserved quantity\footnote{Upon Kaluza-Klein reduction
along the compact circle, ${\cal Q}$ is a conserved charge under the gauge
field obtained from reduction of the two-form potential. 
%We are implicitly taking the compact dimensions to have string size, 
%and we have set the string length equal to 1. 
} 
that measures the number of times that the string goes around the circle
before closing in on itself, or alternatively the number of closed
strings singly-wound on the circle. In general, the winding $n\propto {\cal Q}$
can also be defined, with a different proportionality factor, for dual
brane realizations of the $N=1$ solution, while for
$N=2,3$ obtained from brane
intersections, the winding of the ``effective string" at the
intersection is proportional to ${\cal Q}^N$. 
\begin{figure}[th]
%\hskip1cm
\begin{picture}(0,0)(0,0)
{
\put(34,28){$S^2$}
}
\end{picture}
\centering{\psfig{file=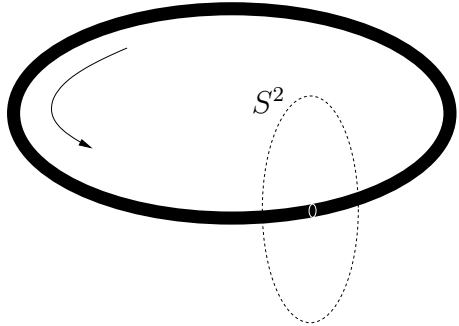,width=6cm}} 
%\begin{center}
%\epsfxsize=10cm\epsfbox{ringdipole.eps}
%\psfig{figure=ringdipole.eps}%
%\end{center}\hskip.5cm
\caption{\small The non-topological winding number $n$ of the ring is
proportional to the local charge ${\cal Q}$ measured from the electric
flux of $H$ across an $S^2$
that encloses a section of the string. An azimuthal angle has been
suppressed in the picture, so
the $S^2$ is represented as a circle.}
\label{fig:ringdipole}
\end{figure}
Consider now a circular string loop in asymptotically flat spacetime (no
compact circle). We can also regard it as a closed string, but now it
does not wrap any topologically non-trivial cycle. Instead, it winds
around a contractible circle, but we can still define the local charge
${\cal Q}$ \reef{elcharge} (see fig.~\ref{fig:ringdipole}), and $n$ as in
\reef{winding}. Now $n$ is not topological, but it still measures the
number of windings of the string around the circle. So even in the case
where the charge is not a conserved quantity, it has a clear physical
meaning and it provides the most natural characterization of the source
of the dipole field. 

Dipole rings are therefore specified by the three physical parameters
$(M,J,{\cal Q})$. The third parameter, which is independent of the other two,
is not a conserved charge, and is classically a continuous parameter.
So, as we will see in detail, it implies infinite non-uniqueness in five
dimensions. Upon quantization in string theory these parameters become
discrete, and there will be a finite but still very large number of
states with the same mass and spin. 

We are now ready to summarize the features of the dipole rings that follow
from the detailed analysis of the next section. We can adequately fix the overall
scale of the solutions by fixing their mass $M$. Then the
solutions are characterized by reduced dimensionless magnitudes, obtained by
dividing out an appropriate power of $M$, or of $G M$ (which has dimension
(length)$^2$), \eg we define a dimensionless
``reduced spin" variable $j$, conveniently normalized as
\beq\label{etadef}
j^2\equiv \frac{27\pi}{32G }\frac{J^2}{M^3}\,,
\eeq
({$j^2$ is often a more
convenient variable than $j$), as well as a reduced area of the horizon,
\beq\label{zetadef}
{a_H}\equiv\frac{3}{16}\sqrt{\frac{3}{\pi}}\frac{{\cal A}_H}{(G M)^{3/2}}\,,
\eeq
and a reduced local charge,
\beq\label{qdef}
q\equiv \frac{{\cal Q}}{\sqrt{G M}}\,.
\eeq

The properties of the neutral solutions ($q=0$) found in \cite{ER} are
summarized in the plot of $a_H$ vs $j^2$ in figure
\ref{fig:ajneu}. The explicit analytical form of
the curves is given below in \reef{zetaeta}, \reef{zetaetabh}. The
non-uniqueness in $27/32\leq j^2<1$ is clear from the
figure. Note we have normalized ${a_H}$ so that its maximum value for
a neutral ring is $1$. 
\begin{figure}[!th]
\label{fig:ajneu}
%\hskip1cm
\begin{picture}(0,0)(0,0)
{\small 
\put(23,48){black hole}
\put(60,18){large black ring}
\put(16,10){small black ring}
}
{\large
\put(60,-9){$j^2$}
\put(-12,30){$a_H$}
}
\small{
\put(0,-4){0}
\put(37,-4){$\frac{27}{32}$}
\put(46,-4){1}
\put(-7,61){$2\sqrt{2}$}
\put(-2,21){$1$}
}
\end{picture}
%\begin{center}
%\epsfxsize=9cm\epsfbox{AJneu.eps}
%\psfig{figure=AJneu.eps}%
%\end{center}\hskip.5cm
\centering{\psfig{file=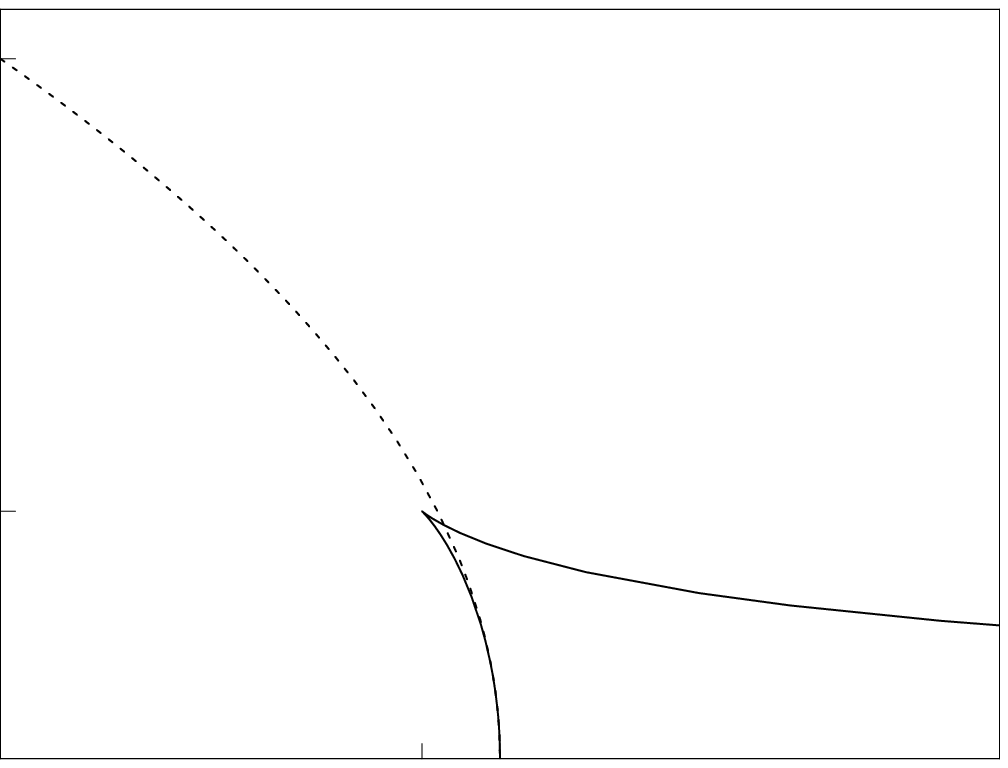,width=9cm}} 
\vspace{4ex}
\caption{\small 
Plot of horizon area vs\ (spin)$^2$, for given mass, for the neutral
rotating black ring (solid) and black hole (dotted). The reduced
variables $j^2$ and $a_H$ are defined in \reef{etadef},
\reef{zetadef}. There are two
branches of black rings, which branch off from the cusp at 
$(j^2,a_H) = (27/32,1)$, and which are dubbed ``large" and ``small"
according to their area. For spins in the range $27/32\leq j^2<1$
the black rings in the two branches coexist with a black hole of the
same mass and spin, implying three-fold non-uniqueness. 
Other interesting features are:
The black hole at $j^2=27/32$, \ie with the
same mass and spin as the minimally spinning ring, has
${a_H}=\sqrt{5}/2$. 
At $j^2={a_H}^2=8/9$ 
% (\ie $\nu=1/3$ in \reef{zetaeta}), 
the curves intersect and we find a black hole and a (large) black ring
both with the same
mass, spin and area. The limiting solution at $(j^2,
a_H)=(1,0)$ is a naked singularity. Fastly spinning black rings, 
$j^2\to\infty$, become thinner and their area decreases as ${a_H}\sim
1/(j\sqrt{2})$.
}
\end{figure}

For the dipole-charged solutions we focus on the values of the dilaton coupling
most relevant to string theory, namely $N=1,2,3$ in \reef{alphan}.
The space of solutions is a two-dimensional surface in the space
$(j,q,{a_H})$. Rather than a 3D plot of this surface, a clearer
representation is obtained plotting several sections of it at constant
$q$ as curves in the plane $(j^2,{a_H})$. These are depicted in
fig.~\ref{fig:ajn}. 
\begin{figure}[!ht]
%\hskip1cm
%\centering{\psfig{file=AJN1.eps,width=7.5cm}\psfig{file=AJN2.eps,width=7.5cm}}
%\vspace{10ex}
%\centering{\psfig{file=AJN3.eps,width=7.5cm}
\begin{center}
\epsfxsize=7.5cm\epsfbox{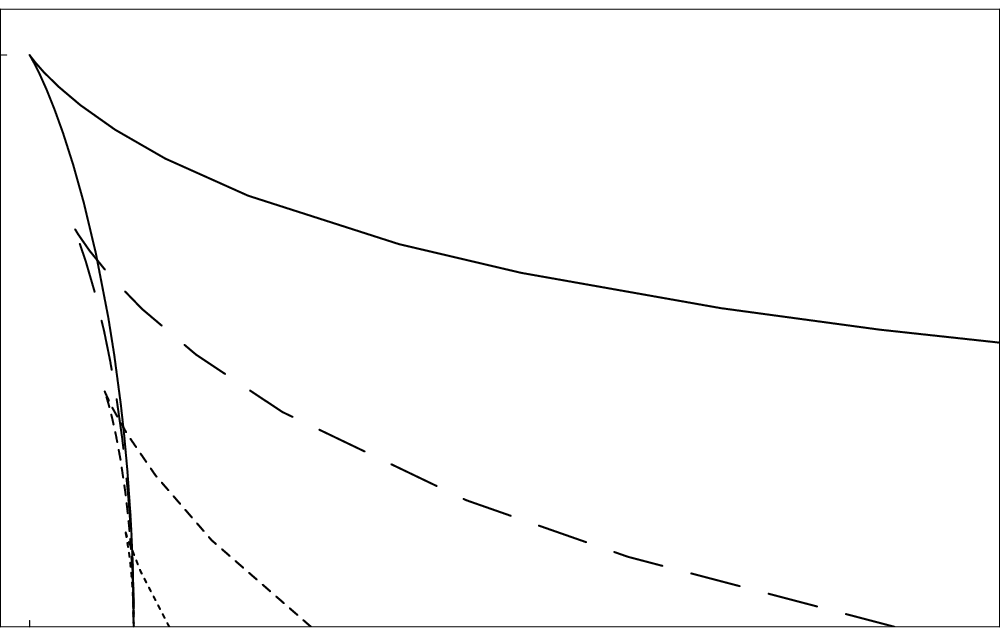}\hskip.5cm
\epsfxsize=7.5cm\epsfbox{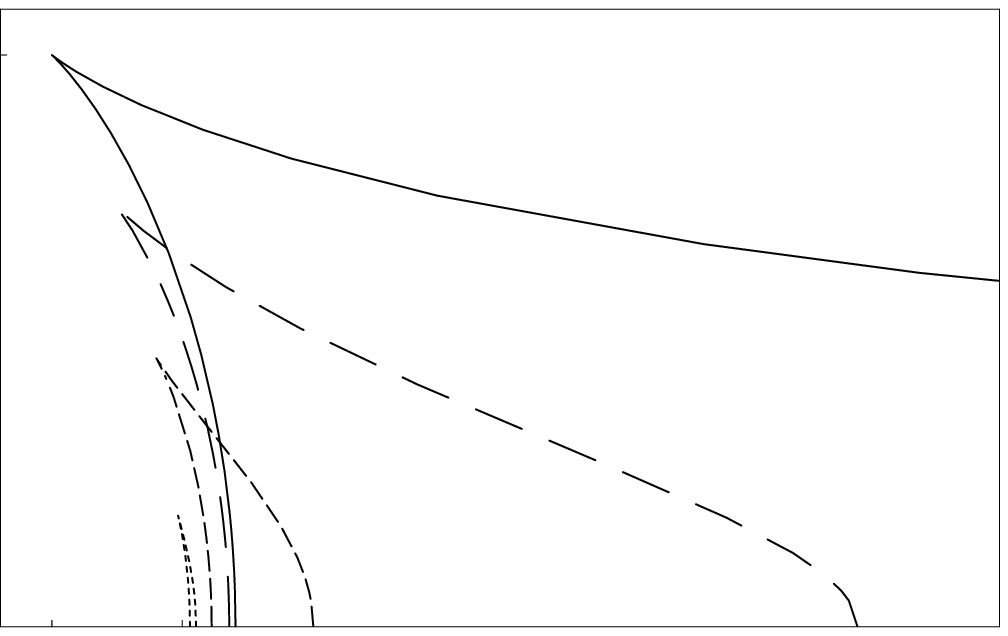}\\
\vspace{10ex}
\epsfxsize=8cm\epsfbox{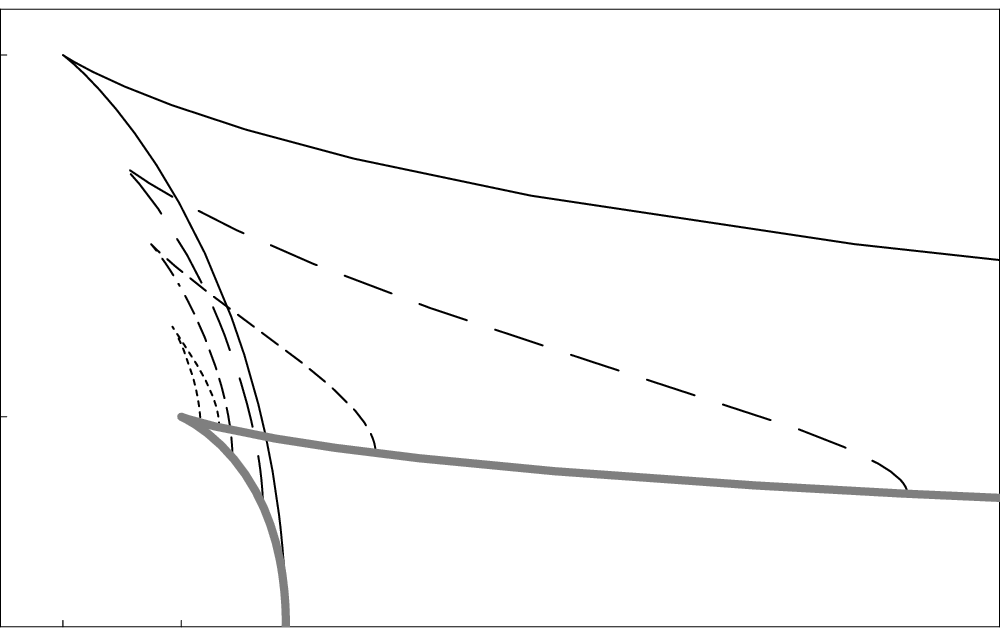}
\begin{picture}(0,0)(120,0)
\put(62,108){$N=1$}
\put(141,108){$N=2$}
\put(102,43){$N=3$}
{\large
\put(90,-10){$j^2$}
\put(25,30){$a_H$}
}
\put(33,83){\vector(-3,-2){10}}
\small{
\put(30,84){$q$}
}
\small{
%\put(0,0){0}
\put(-2,110){1}
\put(-2,68){0}
\put(2,63){$\frac{27}{32}$}
\put(11,64){1}
\put(84,63){$\frac{27}{32}$}
\put(92,64){.95}
\put(99,64){1}
\put(35,45){1}
\put(32,15){.37}
\put(35,0){0}
\put(42,-5){$\frac{27}{32}$}
\put(50,-4){.93}
\put(61,-4){1}
}
\end{picture}
\end{center}\hskip.5cm
\caption{\small $a_H$ vs $j^2$, for different values of $q$,
for dipole rings with dilaton coupling $\alpha=2\sqrt{2/3}, \sqrt{2/3},
0$ ($N=1,2,3$). $q$ varies continuously, but we have plotted the
curves for only a
few representative values, with $q$ increasing in the direction of
the arrow: shorter dashing corresponds to larger $q$, and the solid
curves correspond to the neutral ring ($q=0$).
For any fixed $j$, \ie fixed mass and spin, such that $j^2>27/32$, there
are always rings with $q$ in a continuous range
of values, implying continuous non-uniqueness. Observe also that for
fixed $q\neq 0$, $j$ is bounded above and below so the range
of allowed spins is finite, and becomes narrower as $q$ grows. When $q$
reaches its maximum value the curves degenerate to a point, which is at
$(j^2,a_H)=(1,0)$, $(.95,0)$, $(.93,.37)$ for $N=1,2,3$ resp. (see
fig.~\ref{fig:extr} for the maximum values of $q$). For
$N=3$, the endpoints of the curves lie on the thick grey line, which corresponds to
extremal dipole rings of finite horizon area.
}
\label{fig:ajn}
\end{figure}
It becomes apparent that if we fix the mass and the angular momentum,
with $j^2>27/32$, there exist black ring solutions with $q$
taking values over a continuous finite interval\footnote{For generic
dilaton coupling, the range of $q$ is finite for $1\leq N\leq 3$,
and infinite for $0<N<1$, although these figures do not reflect it.}.
Black hole uniqueness is then violated by a continuous parameter, and
therefore in an infinite manner.

There are other features common to all values of $N$. Like in the
neutral case, for fixed $q$ there are two branches of black rings,
one of them having larger area than the other, and in the range where
they coexist we find two black rings with the same mass, spin and local
charge. The two branches join at the slowest spinning ring with the
given $q$, and this ring always has larger spin and smaller area
than the minimally spinning neutral ring. The smaller area is easy to
understand if we recall that adding charge to a black hole while keeping
its mass fixed typically reduces its area. The larger spin is needed in
order to counterbalance the attraction between diametrically opposite
sections of the ring, which as we saw can be regarded as oppositely
charged.

In contrast to the neutral case, where the spin of large rings is
unbounded, the spin of large dipole rings has an upper bound. The
solutions which saturate the bound are extremal (but non-supersymmetric)
rings, meaning that the outer and inner horizons come to coincide. A
qualitative explanation for this upper bound will be provided
in section \ref{floop}. Of these extremal solutions, the only ones that
have non-zero area are the non-dilatonic extremal rings, \ie $N=3$. This
was expected, since the extremal limit of the straight strings in five
dimensions results in non-zero area only when $N=3$. For $N>1$ the small
ring branch also has an extremal limit, which
corresponds to maximum $j$ in that branch, and which have non-zero area
only if $N=3$. In the $N=3$ plot in fig.~\ref{fig:ajn} we draw the area
of the extremal solutions as a thick grey line, but one should bear in
mind that $q$ is not constant along this curve, \ie the relation between
${\cal Q}$ and $M$ at extremality is not fixed but changes with $J$. 
For $N=1$ the branch of small rings terminates
at a singular solution before the extremal limit is reached.

In section \ref{brane} we exhibit rotating dipole ring solutions of
string and M-theory at low energies, that arise from the intersection of
three kinds of branes with different dipole charges, \eg D2, D6 and NS5
brane charges, and which reduce to the cases $N=1,2,3$ above by having
equal numbers of one, two, or three of the component branes, and no
branes of the remaining components. Their features are qualitatively the same
as described above, according to the number of component branes that are
present.

As we will see, many of the qualitative properties of large rings are
similar to those of their straight string limits. Since we often have a
microscopic stringy picture of the latter, this allows us to get at
least a qualitative microscopic picture of large rings, and understand
some of the features described above, as we will explain in sections
\ref{floop} and \ref{brane}. The small ring branch is more intriguing,
but we will be able to make some reasonable suggestions about its
meaning in section \ref{discuss}.

\medskip

We finish this section mentioning how earlier ring solutions are related
to the ones in this paper. The first example in this class that we are
aware of is the construction of a self-gravitating static loop of string
described in \cite{dggh}. This is a static, extremal dipole ring with
dilaton coupling $N=1$. It does not have a regular horizon, and in the
absence of rotation or any external field, it contains a conical
singularity disk. Ref.~\cite{tubular} generalized it to static extremal
dipole rings with arbitrary dilaton coupling $\alpha$, and also to static
extremal rings from triple intersections of branes. All these solutions
are recovered (in different coordinates) from the solutions in this
paper by setting the rotation to zero and taking the extremal limit.
When $N=3$ they have a regular degenerate horizon, but they still
contain conical singularities (if no fluxbrane is added).
Refs.~\cite{HE,EE} constructed charged black rings, but these have a
\textit{net} charge (besides dipole charges, not independent of the net
charges), and are therefore different from the ones in this paper.

We turn now to the explicit form of the solutions, and to explain how
the features we have described are obtained.

\setcounter{equation}{0}
\section{Dipole black ring solutions}
\labell{dipolerings}

The neutral rotating black ring was obtained in \cite{ER} from the
Kaluza-Klein C-metric solutions in \cite{KKc} via a double Wick rotation
of coordinates and analytic continuation of parameters. All the new
solutions in this paper are similarly obtained from the generalized
C-metrics found in \cite{myc}. It turns out that the coordinates and
parameters in which these new black rings are expressed more simply are
slightly different from those previously used for black rings in
\cite{ER,EE}, so we need to begin with a brief description of the
neutral solution in these new coordinates. They share the feature with
the coordinates in \cite{hongteo,EE} that the cubic functions involved
take a factorized form and allow for easier analytic evaluation than in
the original form in ref.~\cite{ER}. This allows us to provide some new
analytical results.

\subsection{Neutral ring}

The metric is
\beqa\label{neutral}
ds^2&=&-\frac{F(y)}{F(x)}\left(dt+C(\nu,\lambda)\: R\:\frac{1+y}{F(y)}\:
d\psi\right)^2\nonumber\\[2mm]
&&+\frac{R^2}{(x-y)^2}\:F(x)\left[
-\frac{G(y)}{F(y)}d\psi^2-\frac{dy^2}{G(y)}
+\frac{dx^2}{G(x)}+\frac{G(x)}{F(x)}d\varphi^2\right]\,,
\eeqa
where\footnote{We warn the reader that even if we use the same letters
$x,y,\lambda,\nu,F,G$ as in \cite{EE}, their meaning here is slightly
different.}
\beq\label{fandg}
F(\xi)=1+\lambda\xi,\qquad G(\xi)=(1-\xi^2)(1+\nu\xi)\,,
\eeq
and
\beq\label{coeff}
C(\nu,\lambda)=\sqrt{\lambda(\lambda-
\nu)\frac{1+\lambda}{1-\lambda}}\,.
\eeq
The coordinates $x$ and $y$ vary within the ranges 
\beq\label{xyrange}
-1\leq x\leq 1\,,\qquad -
\infty<y\leq -1\,,
\eeq 
and the dimensionless parameters $\lambda$ and $\nu$
within
\beq\label{lanurange}
0< \nu\leq\lambda<1\,.
\eeq 
$R$ has dimensions of length, and for thin large rings it corresponds
roughly to the radius of the ring circle \cite{EE}. In order
to avoid conical
singularities at $y=-1$ and $x=- 1$ the angular variables must be
identified with
periodicity
\beq\label{period0}
\Delta\psi=\Delta\varphi=2\pi\frac{\sqrt{1-\lambda}}{1-\nu}\,.
\eeq
To avoid also a conical singularity at $x=+1$ we take the
two parameters $\lambda$, $\nu$, to be related as
\beq\label{equil0}
\lambda=\frac{2\nu}{1+\nu^2}\,.
\eeq
With these choices, the solution has a regular horizon of topology
$S^1\times S^2$ at $y=-1/\nu$, an ergosurface of the same topology at
$y=-1/\lambda$, and an inner spacelike singularity at $y=-\infty$.
Asymptotic spatial infinity is reached as $x\to y\to -1$. The coordinate
system is illustrated in fig.~\ref{fig:coords}.
\begin{figure}[!th]
\begin{picture}(0,0)(0,0)
\footnotesize{
%\put(0,0){0}
\put(13,34){$x=-1$}
\put(49,34){$x=+1$}
\put(33,40){$x$}
\put(59,64){$\psi$}
\put(83,63){$y$}
\put(84,40){$y=-1/\nu$}
\put(53,6){$y=-1$}
\put(38,-3){$x=\mathrm{const}$}
}
\end{picture}
\centering{\epsfxsize=11cm\epsfbox{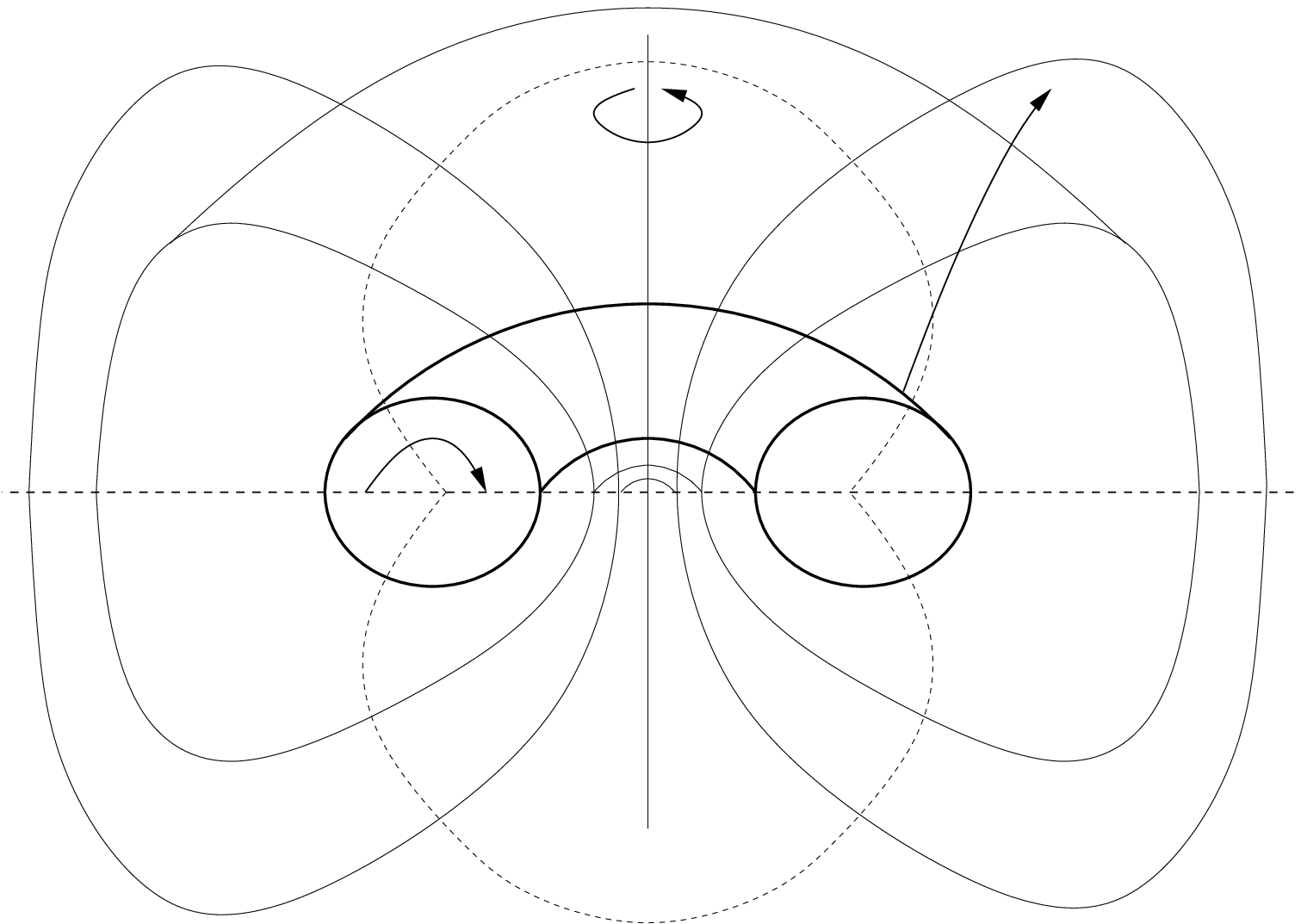}}
\vskip.5cm
\caption{\small Coordinate system for black ring metrics (adapted from
\cite{ER2}). The diagram sketches a section at constant $t$ and
$\varphi$. Surfaces of constant $y$ are ring-shaped, while $x$ is a
polar coordinate on the $S^2$ (roughly $x\sim\cos\theta$). $x=\pm 1$ and
$y=-1$ are fixed-point sets (\ie axes) of $\partial_\varphi$ and
$\partial_\psi$, resp. Infinity lies at $x=y=-1$.
}
\label{fig:coords}
\end{figure}

A static ring solution, which necessarily contains a conical
singularity, is obtained when instead of \reef{equil0} we set
$\lambda=\nu$ \cite{ER2}. The spherical black hole of Myers and Perry
with rotation in a single plane can also be obtained from
\reef{neutral}, but in the coordinates in \reef{neutral} this case is much
subtler than when using the forms of the metric employed in
\cite{ER,EE}. The appropriate limiting procedure is described in the
appendix. When $\nu\to 1$ the solution becomes singular and the horizon
is replaced by a naked singularity.

The mass, spin, and other physical parameters of the neutral black
ring can be readily recovered as the neutral limit $\mu\to 0$ of the
expressions for the dipole ring,
eqs.~\reef{massN}-\reef{omegaN} below,
so we will not present them here. Note that for a black ring
at equilibrium, \ie satisfying \reef{period0} and \reef{equil0}, there
is only one independent dimensionless parameter, so the reduced area and
spin, \reef{etadef}, \reef{zetadef}, must be related. ${a_H}$ and
$j$ can be computed from the expressions
for $M$, $J$ and ${\cal A}_H$, and the analytic relation between them
is, in parametric form,
\beq\label{zetaeta}
{a_H}= 2\sqrt{\nu(1-\nu)}\,,\qquad
j^2= \frac{(1+\nu)^3}{8\nu}\qquad \mathrm{(black\: ring)}\,,
\eeq
with $0<\nu\leq 1$. On the other hand, for the spherical black hole, 
\beq\labell{zetaetabh}
{a_H}=2\sqrt{2(1-j^2)}\qquad \mathrm{(black\: hole)}\,.
\eeq 
These curves were plotted and discussed in figure \ref{fig:ajneu}.

Finally, in order to see how we recover a boosted straight black string
in the limit where the radius of the ring circle grows very
large, define
\beq\label{limit0}
r_0=\nu R\,,\qquad \cosh^2\sigma=\frac{\lambda}{\nu}\,,
\eeq
and
\beq\label{limit0c}
r=-\frac{R}{y}\,,\quad \cos\theta=x\,,\quad w= R\psi
\eeq
and take the limit $R\to \infty$, $\lambda,\nu\to 0$, keeping
$r_0,\sigma$ and $r,\theta,w$
finite. Then we obtain the
metric for a boosted black string
\beq\label{straight0}
ds^2=-\hat f\left(dt-\frac{r_0\sinh\sigma\cosh\sigma}{r\hat
f}\:dw\right)^2 +\frac{f}{\hat f}\:dw^2+\frac{dr^2}{f}+r^2d\Omega_{(2)}^2
\eeq
where
\beq\label{ffs}
f=1-\frac{r_0}{r}\,,\qquad \hat f=1-\frac{r_0\cosh^2\sigma}{r}\,.
\eeq
The value of the boost obtained from the limit of rings at equilibrium
is, from \reef{equil0} and \reef{limit0}, $|\sinh\sigma|=1$, and this
was shown in \cite{EE} to make the ADM pressure $T_{ww}=0$ (see
eq.~\reef{setensor} below, with $\gamma=0$) . 

Even if $\lambda$ and $\nu$ do not directly
correspond to physical quantities (whereas $j^2$ and ${a_H}$
do), eq.~\reef{limit0} allows to find an approximate meaning for
them, which becomes more accurate for thin rings. For large but finite $R$, the
parameter $\nu\simeq r_0/R$ measures the ratio between the radius of the
$S^2$ at the horizon and the radius of the ring. So smaller values of
$\nu$ correspond to thinner rings. Also, $\lambda/\nu$ is a measure of
the speed of rotation of the ring. More precisely,
$\sqrt{1-(\nu/\lambda)}$ can be approximately identified with the local
boost velocity $\tanh\sigma$, which for rings in equilibrium,
\reef{equil0}, is $\sqrt{(1-\nu^2)/2}$.

\subsection{Dipole rings}
\label{dipolsol}

The dipole black ring solutions contain a new parameter $\mu$, related
to the local charge ${\cal Q}$ of the ring. For arbitrary dilaton coupling
$\alpha$, expressed in terms of $N$ as in \reef{alphan}, the geometry of
the solutions is 
\beqa\label{magnetic}
ds^2&=&-\frac{F(y)}{F(x)}\left(\frac{H(x)}{H(y)}\right)^{N/3}
\left(dt+C(\nu,\lambda)\: R\:\frac{1+y}{F(y)}\: d\psi\right)^2\\[3mm]
&&+\frac{R^2}{(x-y)^2}\: F(x)\left(H(x)H(y)^2\right)^{N/3}\left[
-\frac{G(y)}{F(y)H(y)^N}d\psi^2-\frac{dy^2}{G(y)}
+\frac{dx^2}{G(x)}+\frac{G(x)}{F(x)H(x)^N}d\varphi^2\right]\,.\nonumber
\eeqa 
The functions $F$ and $G$ are like in \reef{fandg}, and 
\beq
H(\xi)=1-\mu\xi\,. 
\eeq 
If we consider the magnetic solutions of the
Einstein-Maxwell-dilaton theory \reef{e2daction} then the gauge
potential is 
\beq
A_\varphi=C(\nu,-\mu)\:\sqrt{N}\:R\:\frac{1+x}{H(x)}+k_1\,, 
\eeq 
(the constant $k_1$ moves Dirac strings around) and the dilaton 
\beq
e^{-\phi}=\left(\frac{H(x)}{H(y)}\right)^{N\alpha/2}\,. 
\eeq
$C(\nu,-\mu)$ is like in \reef{coeff} but with $\lambda\to -\mu$. Since
this field is purely magnetic, it makes no contribution to the 
Chern-Simons term required by five-dimensional supergravity, and so
the solution belongs in this theory as well.

For the
electric solutions of \reef{e2daction}, the metric is the same as above,
the dilaton $\tilde\phi=-\phi$, and the two-form potential 
\beq
B_{t\psi}=C(\nu,-\mu)\:\sqrt{N}\:R\:\frac{1+y}{H(y)}+k_2\,. 
\eeq 
The
constant $k_2$ may be chosen at convenience, \eg to make $B_{t\psi}$
vanish at $y=-1/\nu$. 
%\beq
%B_{t\psi}=\sqrt{\frac{\mu(1-\mu^2)}{\mu+\nu}}\:\sqrt{N}\:R\:\frac{F(y)}{H(y)}\,. 
%\eeq

The parameters $\lambda$ and $\nu$ vary in the same ranges as in the neutral case
\reef{lanurange}, while
\beq\label{murange}
0\leq \mu<1\,.
\eeq
When $\mu=0$ we recover the neutral solution \reef{neutral}.

Most of the features of the solutions are analyzed in the same manner as
for the neutral black ring, so we refer to the previous section and
earlier papers \cite{ER}, \cite{EE} for more details. The coordinate $x$
varies in $[-1,1]$. Initially we take $y\in (-\infty, -1]$, but we will
shortly see that this can be extended across $|y|=\infty$ to the range
$(1/\mu,+\infty)$. The possible conical singularities at the axes
extending to infinity, $x=-1$ and $y=-1$, are avoided by setting
\beq\label{periodN}
\Delta\psi=\Delta\varphi=4\pi\frac{H(-1)^{N/2}\sqrt{F(-1)}}{|G'(-1)|}=
2\pi\frac{(1+\mu)^{N/2}\sqrt{1-\lambda}}{1-\nu}\,.
\eeq
The balance between forces in the ring will be achieved when, in
addition, there are no conical singularities at $x=+1$. This requires
that
\beq
\Delta\varphi =4\pi\frac{H(+1)^{N/2}\sqrt{F(+1)}}{|G'(+1)|}\,,
\eeq
which can be satisfied simultaneously with \reef{periodN} only if
\beq\label{equilN}
\frac{1-\lambda}{1+\lambda}\left(\frac{1+\mu}{1-\mu}\right)^N=
\left(\frac{1-\nu}{1+\nu}\right)^2\,.
\eeq
In the neutral case $\mu=0$ this equation is solved by \reef{equil0}. 
%The rotation disappears when
%$\lambda=\nu$. In this case it
%is impossible to satisfy the equilibrium condition \reef{equilN}.

Under these conditions, it is easy to see that the solution has a regular outer
horizon of topology $S^1\times S^2$ at $y=-1/\nu$. In addition, there is
an inner horizon at $y=-\infty$. The metric can be continued beyond this
horizon (as in \eg \cite{ER}) to positive values $1/\mu<y<\infty$, until
$y=1/\mu$ is hit from above, which is a curvature singularity. The two
horizons coincide when $\nu=0$, which defines the extremal limit, and $\nu$
can be regarded as a non-extremality parameter. In general,
a ring-shaped ergosurface is present at $y=-1/\lambda$.

The calculation of the mass, angular momentum, horizon area, temperature
(from surface gravity) and angular velocity at the (outer) horizon is
straightforward, and one finds
\beq\label{massN}
M=\frac{3\pi R^2}{4G }\frac{(1+\mu)^N}{1-
\nu}\left(\lambda+\frac{N}{3}\frac{\mu(1-\lambda)}{1+\mu}\right)\,,
\eeq

\beq\label{spinN}
J=\frac{\pi R^3}{2G }\frac{(1+\mu)^{3N/2}\sqrt{\lambda(\lambda-
\nu)(1+\lambda)}}{(1-\nu)^2}\,,
\eeq

\beq\label{areaN}
{{\cal A}_H}=8\pi^2 R^3
\frac{(1+\mu)^N\nu^{(3-N)/2}(\mu+\nu)^{N/2}\sqrt{\lambda(1-
\lambda^2)}}{(1-\nu)^2(1+\nu)}\,,
\eeq

\beq\label{tempN}
T=\frac{1}{4\pi
R}\frac{\nu^{(N-1)/2}(1+\nu)}{(\mu+\nu)^{N/2}}\sqrt{\frac{1-
\lambda}{\lambda(1+\lambda)}}\,,
\eeq

\beq\label{omegaN}
\Omega=\frac{1}{R}\frac{1}{(1+\mu)^{N/2}}\sqrt{\frac{\lambda-
\nu}{\lambda(1+\lambda)}}\,.\\[2mm]
\eeq
We can also compute the local charge ${\cal Q}$ defined in
\reef{elcharge}. The integral is taken over an $S^2$ parametrized
by $(x,\varphi)$, at constant $t$, $\psi$ and $y\in(-1/\nu,-1)$, see
fig.~\ref{fig:coords}.
Then
\beq\label{QN}
{\cal Q}=R\sqrt{N}\frac{(1+\mu)^{(N-
1)/2}\sqrt{\mu(\mu+\nu)(1-\lambda)}}{(1-\nu)\sqrt{1-\mu}}\,.
\eeq
In addition, we define the potential $\Phi$ from the difference between
the values of $B$ at infinity and at the horizon, 
\beq
\Phi=\frac{\pi}{2G }\left[B_{t\tilde\psi}(x=y=-1)-B_{t\tilde\psi}(y=-1/\nu)\right],
\eeq
where $\tilde\psi=(2\pi/\Delta\psi)\psi$ is the canonically normalized
angular variable, and the factor $\pi/2G $ is introduced simply for
convenience. Then
\beq\label{phiN}
\Phi=\frac{\pi R}{2G }\sqrt{N}\frac{(1+\mu)^{(N-1)/2}\sqrt{\mu(1-\mu)(1-
\lambda)}}{\sqrt{\mu+\nu}}\,.
\eeq

A straightforward calculation using these results shows that the
black ring satisfies a Smarr
relation\footnote{
The equilibrium condition \reef{equilN} has not been enforced in any of
the expressions \reef{massN}-\reef{smarr}, so in principle they are valid as well for
unbalanced black rings, although we will not be considering these in any
detail in this paper.}
\beq\label{smarr}
M=\frac{3}{2}\left(\frac{1}{4G }{{\cal A}_H}T+\Omega J\right)+\frac{1}{2}{\cal Q}\Phi\,.
\eeq
The first law
\beq\label{firstlaw}
dM=\frac{1}{4G }Td{{\cal A}_H} +\Omega dJ+\Phi d{\cal Q}\,,
\eeq
can also be, somewhat laboriously, verified explicitly. The numerical
coefficients in \reef{smarr} and \reef{firstlaw} are consistent with the
homogeneity properties of $M$ as a function 
of scaling dimension 2, of the variables ${\cal A}_H$, $J$, ${\cal Q}$
with scaling dimensions 3, 3, 1, resp.

The horizon area of dilatonic rings vanishes in the extremal limit
$\nu\to 0$. These are nakedly singular solutions. Only in the
non-dilatonic case $N=3$ does the area remain finite as $\nu\to 0$,
and the solution has a regular degenerate horizon. The mass $M$ and
local charge ${\cal Q}$ are not any simply related in the extremal limit, and for
finite values of $R$, the extremal solutions are not supersymmetric. 

The main properties of dipole rings are summarized in
fig.~\ref{fig:ajn}. In order to produce these plots, we first solve
eq.~\reef{equilN} for $\lambda$ as a function of $\nu$ and $\mu$. Using
\reef{massN}-\reef{QN}, we compute the reduced local charge $q$, spin
$j$, and area $a_H$, as functions of $\mu$ and $\nu$, and then
we invert to find $\mu$ as a function of $q$ and $\nu$. The
inverted function can be seen to involve more than one branch for $N>1$.
All of this can be done fully analitically for $N=1,2,3$, but the
expressions are exceedingly long and only simplify in the extremal
limit. So we proceed using a symbolic manipulation computer program.
Eventually we obtain $a_H$ and $j^2$ as functions of $\nu$ and
$q$, which allows us to plot the curves for fixed $q$, varying
$0\leq\nu\leq 1$. We refer to section \ref{setup}
for a discussion.

The limit where the ring becomes a straight string allows us again to
get a feeling for other properties of the solutions. In addition to
taking $\lambda,\nu\to 0$, $R\to\infty$ with \reef{limit0} and
\reef{limit0c} finite, we also take $\mu\to 0$ and keep finite
\beq
\mu R=r_0\sinh^2\gamma\,,
\eeq
where $\gamma$ gives a convenient parametrization of the charge in this
limit. Then
the limiting solution is
\beq\label{straightN}
ds^2=-\frac{\hat f}{h^{N/3}}\left(dt-\frac{r_0\sinh\sigma\cosh\sigma}{r\hat
f}\:dw\right)^2 
+\frac{f}{h^{N/3}\hat f}\:dw^2+h^{2N/3}\left(\frac{dr^2}{f}+r^2d\Omega_{(2)}^2\right)\,,
\eeq
\beq
A_\varphi=\sqrt{N}r_0\sinh\gamma \cosh\gamma(\cos\theta+1)\,,
\qquad e^{\phi}=h^{N\alpha/2}\,,
\eeq
for the magnetic solution, and
\beq
B_{tw}=\sqrt{N}r_0\sinh\gamma \cosh\gamma (h^{-1}-1)\,,
\qquad e^{\tilde\phi}=h^{-N\alpha/2}\,,
\eeq
for the electric one,
where $f$ and $\hat f$ were defined in \reef{ffs}, and
\beq\label{hs}
h=1+\frac{r_0\sinh^2\gamma}{r}\,.
\eeq
These are five-dimensional charged strings with a momentum wave.
When $N=3$ and $\gamma=\sigma$, the reduction along $w$ to four dimensions
yields the Reissner-Nordstrom black hole. 

While the straight string \reef{straightN} solves the field equations
for arbitrary values of $\sigma$ and $\gamma$, the strings that are
obtained as a limit of rings that satisfy the
equilibrium condition \reef{equilN} must, from this condition, be such that
\beq\label{equilNs}
\sinh^2\sigma=1+N\sinh^2\gamma\,.
\eeq
Note that when $\gamma\neq 0$ the boost
has to be larger than in the neutral case $\gamma=0$. This is easily interpreted:
sections of the ring at diametrically opposite ends, $\psi$ and
$\psi+\pi$, have opposite orientation and therefore they attract each
other via the $H_{\mu\nu\rho}$ field. Then a larger centrifugal repulsion
is needed in order to achieve equilibrium, and the effect persists even
at very large ring radii.

The ADM stress-energy tensor of the string
\reef{straightN} is
\beqa\label{setensor}
T_{tt}&=&\frac{r_0}{4G }(1+\cosh^2\sigma+N\sinh^2\gamma)\,,\nonumber\\
T_{ww}&=&\frac{r_0}{4G }\left(\sinh^2\sigma-1-N\sinh^2\gamma\right)\,,\\
T_{tw}&=&\frac{r_0}{4G }\sinh\sigma\cosh\sigma\,.\nonumber
\eeqa
We observe that, as was the case for the neutral ring and the charged rings
of \cite{EE}, the pressure $T_{ww}$ of the
limiting string vanishes when the equilibrium condition
\reef{equilNs} is enforced. The restrictions on the parameters of the
strings that result as a limit of balanced rings, such as
eq.~\reef{equilNs}, are in general
poorly understood (see also \cite{EE}), and although it is clear that they are the
result of an equilibrium of forces, it would be interesting to
understand them in more detail.
At any rate, in the limit we still recover the
identification
\beq
\frac{M}{2\pi R}\to T_{tt}\,,\qquad \frac{J}{2\pi R^2}\to T_{tw}\,,
\eeq
while ${\cal Q}\to \sqrt{N}r_0\sinh\gamma\cosh\gamma$ is
identified as the black
string charge under the $H$ field.

\subsection{Extremal rings}
\labell{extremal}

The extremal solutions are defined as the limiting case where the outer
and inner horizons coincide. For the dilatonic solutions the
horizon turns into a null singularity, whereas the non-dilatonic extremal ring
has a regular degenerate horizon. Typically, the extremal limit yields
solutions where the spin and/or the charge reach a maximum value. In the
present case, if we fix the mass and the spin, with $j^2>1$, then there
is always a maximum value of ${\cal Q}$ which is saturated precisely in
the extremal limit $\nu\to 0$. For $27/32<j^2<1$ the situation is more
complicated and depends on the dilaton coupling, as can be seen in
fig.~\ref{fig:ajn}. On the other hand, if we fix the local charge and
the mass, \ie fix $q$, the spin reaches an absolute maximum at an
extremal solution along the large ring branch. There is also a
local maximum along the small ring branch, but this solution is extremal
only for $N>1$. For $N=1$ this local maximum
is not an extremal solution. In this case the branch of small black
rings with $q<q_{\rm max}=\sqrt{2/3\pi}$ terminates at
$\nu=[1-(q/q_{\rm max})^2]/[1+(q/q_{\rm max})^2]$, \ie before $\nu=0$, so
these should not be interpreted as extremal solutions, except when $q\to
q_{\rm max}$.

In the extremal limit the physical variables
admit simple enough expressions in terms of the only dimensionless
parameter that remains after imposing the equilibrium condition
\reef{equilN}, and which we take to be $\mu$. Namely, for the extremal
$N=1$ solution (for which the equilibrium condition becomes simply
$\lambda=\mu$) we find\footnote{Exact expressions for general $N$ can be
found, but they are
fairly long.}
\beq 
j^2=\frac{27(1+\mu)^4}{16\mu(2+\mu)^3}\,,\qquad
{q}=\sqrt{\frac{2\mu}{\pi(2+\mu)}}\,,\qquad
{a_H}=0\,.
\eeq
For $N=2$:
\beq
j^2=\frac{27(1+\mu)^5}{4\mu(4+\mu+\mu^2)^3}\,,\qquad
{q}=\sqrt{\frac{4\mu(1-\mu)}{\pi(4+\mu+\mu^2)}}\,,\qquad
{a_H}=0\,.
\eeq
For $N=3$:
\beq
j^2=\frac{(1+\mu)^6(3+\mu^2)^2}{128\mu(1+\mu^2)^3}\,,\quad
{q}=\sqrt{\frac{\mu(1-\mu)^2}{\pi(1+\mu^2)}}\,,\quad
{a_H}=\sqrt{\frac{\mu(1-
\mu^2)^3(3+\mu^2)}{4(1+\mu^2)^3}}\,.
\eeq

\begin{figure}[!ht]
%\hskip1cm
\begin{picture}(0,0)(0,0)
{\small 
\put(40,33){$N=2$}
\put(60,33){$N=1$}
\put(16,33){$N=3$}
}
{\large
\put(55,-11){$q$}
\put(-12,30){$j^2$}
}
\small{
\put(1,-4){0}
\put(36,-4){$.21$}
\put(46,-4){$.26$}
\put(80,-5){$\sqrt{\frac{2}{3\pi}}$}
\put(-2,16){$1$}
\put(-5,12){$.95$}
\put(-5,8){$.93$}
}
\end{picture}
%\begin{center}
%\epsfxsize=9cm\epsfbox{AJneu.eps}
%\psfig{figure=AJneu.eps}%
%\end{center}\hskip.5cm
\centering{\psfig{file=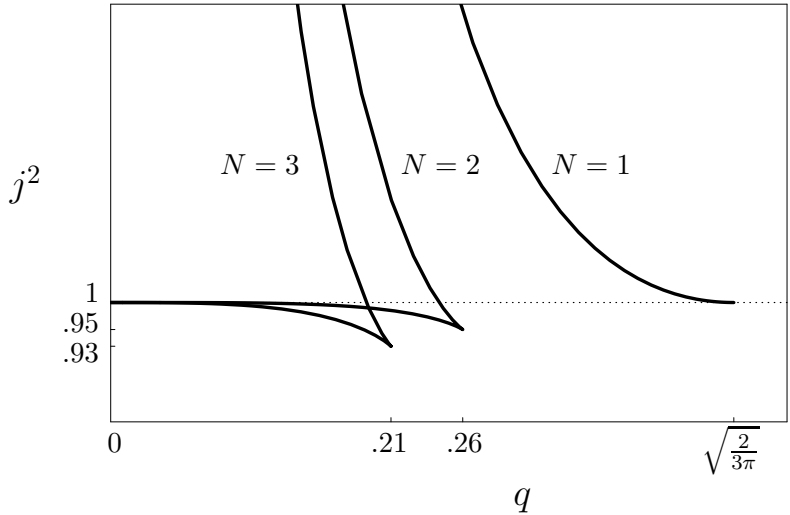,width=9cm}} 
\vspace{9ex}
\caption{\small (Spin)$^2$ vs local charge, for extremal rings of fixed mass.
Although we only plot the cases of integer $N$, the curves for all
$1<N\leq 3$ are qualitatively similar to $N=2,3$, while for $0<N<1$
there is no upper bound for $q$, and the curve asymptotes to $j^2=1$ as
$q\to \infty$.
}
\label{fig:extr}
\end{figure}
We plot $j^2$ vs $q$ in fig.~\ref{fig:extr}. Observe the existence of a
maximum value of $q$ in all three cases (it is also possible to see that
if $N<1$ there is no such upper bound). We will interpret this feature
from a microscopic viewpoint in the next section. For $N=2,3$, when $q$
is below this maximum value there exist two extremal ring solutions with
the same mass and dipole charge, but different spin. These two rings are
the extremal limits of solutions in the large and small ring branches
that we mentioned above. Note the absence of a second $N=1$ extremal
solution, for fixed $q$, in fig.~\ref{fig:extr}.

In \cite{tubular} it was observed that when the $N=3$ extremal static
ring is balanced by a fluxbrane of appropriate strength, then the
geometry near the horizon of the ring becomes exactly AdS$_3\times S^2$.
Ref.~\cite{EE} conjectured that this might be the case as well when the
ring is balanced not by a fluxbrane but by rotation, as in the present
paper. However, closer examination reveals that this is not the case:
for any finite value of $R$ the near-horizon extremal geometry is
distorted away from AdS$_3\times S^2$ by factors of the polar angle of
the $S^2$.

\section{Loop of fundamental string, and qualitative microscopics} 
\labell{floop}

The extremal ring solution is particularly simple when $N=1$. 
%The equilibrium condition \reef{equilN} reduces to $\lambda=\mu$, and
%the metric and the expressions for physical parameters
%\reef{massN}-\reef{phiN} greatly simplify (\eg ${\cal Q}=R\mu$). 
Since the local charge in this case can be regarded as fundamental
string charge, it is expected to describe a rotating loop of fundamental
string. However, this appears to raise a puzzle, since it is well known
that in perturbative string theory it is impossible to have a closed
string loop that rotates like a rigid wheel. The resolution is fairly
simple, and can be understood more simply if we focus first on the
similar situation posed by a straight string that carries linear
momentum. 

Refs.~\cite{cetal,hetal} constructed solutions of low energy string theory with
non-zero winding and momentum of the form
\beq\label{correct}
ds^2=-\frac{1}{h}du\:dv-(h^{-1}-1)|\dot
\mathbf{F}|^2 dv^2+2(h^{-1}-1)\dot\mathbf{F}\cdot d\mathbf{x}\:dv+
d\mathbf{x}\cdot d\mathbf{x}
\eeq
where $u,v$ are light-cone coordinates, $\mathbf{F}(v)$ parametrizes
a curve in the $d-2$ transverse
directions $\bf x$, and $h$ is a harmonic function
\beq
h=1+\frac{Q}{|\mathbf{x}-\mathbf{F}|^{d-4}}
\eeq
(we will not need the dilaton and $H$ field for our argument) so the
solution is singular at $\mathbf{x}=\mathbf{F}(v)$. $Q$ is proportional
to the winding number of the string. It was shown in \cite{hetal} that one
can match these solutions to a fundamental string source, with a profile
of momentum-carrying transverse oscillations given by $\mathbf{F}(v)$.

Now consider fundamental strings with small oscillation amplitudes, such
that when we perform a
coarse-grain average over their oscillation profiles we get $\langle
\mathbf{F}\rangle\simeq 0$ and $\langle \dot \mathbf{F}\cdot
\mathbf{x}\rangle\simeq 0$, but $\langle |\dot\mathbf{F}|^2\rangle\neq 0$. We
shall not be too precise about how this averaging is performed, but one
can suppose, along the lines of \cite{mathur1}, that we choose not to
resolve distances smaller than a typical amplitude, so we remain
at $|\mathbf{x}| \gg |\mathbf{F}|$. The coarse-grained metric is then
approximated by
\beq\label{naive}
ds^2=-\frac{1}{h}du\:dv +\frac{p}{|\mathbf{x}|^{d-4}h} dv^2+
d\mathbf{x}\cdot d\mathbf{x}
\eeq
with $h=1+Q/|\mathbf{x}|^{d-4}$ and $p=Q\langle
|\dot\mathbf{F}|^2\rangle$, \ie we
can regard the strings as oscillating about a center-of-mass line at
$\mathbf{x}=0$, and carrying a large, macroscopic total momentum
density
$Q\langle |\dot\mathbf{F}|^2\rangle$.

So, after averaging over strings with small oscillations, the effective
solution \reef{naive} looks like a
string with a longitudinal momentum wave. Since it is well known that
relativistic fundamental strings cannot support longitudinal
oscillations, the proper interpretation of \reef{naive} must be in terms
of this averaging procedure. This interpretation is, for the purposes of
this paper, basically the same as the microscopic proposal for the two-
and three-charge black hole solutions in \cite{mathur1,mathur2,mathur3}.
Following \cite{mathur2,mathur3}, we may refer to \reef{naive} as the ``naive''
solution for a string carrying momentum, which is a coarse-grained
average over the structure near the core of the ``correct'' metrics
\reef{correct}.

The extremal rings of this paper are now seen to be ``naive" metrics, as
\reef{naive}. Indeed, the limit of infinite radius \reef{straightN} for
the $N=1$ extremal solutions is precisely the metric \reef{naive} in
$d=5$ with $Q={\cal Q}$. So the rotating loop of string is an effective
coarse-grained description of strings which oscillate about the ring,
with amplitude much smaller than the ring radius, and the rotation is
provided by the momentum circulating along the ring carried by these
oscillations. One would also expect the existence of ``correct'' ring
solutions, analogous to \reef{correct}, where the momentum wave is
resolved into transverse oscillations of the string, but due to the
absence of supersymmetry it appears difficult to find these solutions
analytically\footnote{The resolution of donut-shaped configurations in
\cite{mathur1} is different, since those carry a net charge, like
supertubes.}. It would also be interesting to construct the rotating
loops of string, with oscillations, as classical solutions of
perturbative string theory.

This picture allows us to get a qualitative microscopic
understanding of some features of extremal dipole rings. Our main
assumptions are that (a) extremal solutions correspond to strings with
chiral oscillations, \ie only left-movers (or only right-movers), and
(b) that ${\cal Q}$ gives, through eq.~\reef{winding}, the number $n$ of
strings wound in the loop (or alternatively, the number of windings of a
single string $n$ times longer). Then we can see that there must be an
upper bound on the spin for fixed ${\cal Q}$ and $M$, which is saturated
in the extremal limit. For, if we fix $n$ and the total energy $M$, then
we are restricting the total energy that can be carried by the
oscillations. If this energy is distributed among only left-movers, \ie
an extremal state, then the angular momentum will be maximized.
Solutions with both left and right movers will be non-extremal, and have
lower angular momentum. This is indeed the behavior that we observe in
the ring solutions. 

We can also understand qualitatively why $q$ must be bounded above, \ie
why there is a maximum ${\cal Q}$ for fixed $M$.\footnote{This is
non-trivial: for dilaton coupling $N<1$ there is no such a maximum,
suggesting that these cases may not have a microscopic stringy
interpretation.} The total energy of the configuration, $M$,
must be distributed among the tension of wound strings, and among the
momentum carried by oscillations. So there will be a maximum number of
strings that can make a ring (and a similar story goes for a single long
string that winds $n$ times around the loop). The energy that goes into
each string in the loop is roughly the product of the string tension
times the length of the loop, so we would expect to maximize the number
of strings by decreasing the radius. But a ring rotating at a smaller
radius must be rotating faster in order to be balanced, and therefore
more energy has to be spent into momentum carriers. So a compromise must
be reached between momentum and winding, \ie between $j$ and $q$. It is
then fairly clear that, among extremal rings, the solution with maximum
$q$ will also be the one with minimum $j$, and again this is observed in
fig.~\ref{fig:extr}. It is less clear, and this requires a better
understanding of the balance of forces in rings, why the equilibrium
condition \reef{equilN} for the fundamental string loop, $\lambda=\mu$,
corresponds, at large radii, to having equal winding and momentum
numbers, \ie self-T-dual strings.

In section \ref{discuss} we will take these arguments further
to try to understand the qualitative properties of black rings far from
extremality, even neutral rings.

The degeneracy of states of a string with chiral momentum can be
reproduced in the supergravity context by assigning a stretched horizon
to \reef{naive} with entropy proportional to its area \cite{stretch}
(see also \cite{mathur1}). One may try to extend this picture to the
extremal $N=1$ ring, including the corrections for large but finite
radius. However, a more stringent test of the correspondence between
area and entropy of string states is provided by extremal rings with
non-zero area, and these will be analyzed in the next section.

\setcounter{equation}{0}
\section{Rings from brane intersections, and quantitative microscopics}
\labell{brane}

The C-metric solutions of ref.~\cite{myc} can also be used to obtain, via
appropriate analytic continuations, black rings with dipoles under
several gauge fields. These are naturally interpreted in string theory
as arising from intersections of branes. The rotating ring solution with
three local charges is particularly interesting. It can be regarded as
the result of bending into a circular ring-shape a three-charge
five-dimensional black string with a momentum wave. The three charges
can be interpreted as \eg charges of D6, D2 and NS5 branes intersecting
along the string, and the momentum runs along the intersection. In
another realization the ring results from the intersection of three M5
branes \cite{kt}. Since they are in any case related by dualities, we
will provide explicit results only for the more symmetric
M5$\perp$M5$\perp$M5 configuration.

\subsection{Supergravity solution}

The solution contains three new parameters $\mu_i$, $i=1,2,3$, in addition
to $\lambda$ and $\nu$, and which are associated with each of the three
M5 branes. Accordingly, we introduce
\beq
H_i(\xi)=1-\mu_i \xi\,.
\eeq

The full eleven-dimensional metric is
\beqa
ds^2= ds^2_{(5)}
&+&
\left[\frac{H_2(x)H_3(x)}{H_2(y)H_3(y)}\right]^{1/3}
\left(\frac{H_1(y)}{H_1(x)}\right)^{2/3}(dy_1^2+dy_2^2)\nonumber\\
&+&
\left[\frac{H_1(x)H_3(x)}{H_1(y)H_3(y)}\right]^{1/3}
\left(\frac{H_2(y)}{H_2(x)}\right)^{2/3}(dy_3^2+dy_4^2)\\
&+&
\left[\frac{H_1(x)H_2(x)}{H_1(y)H_2(y)}\right]^{1/3}
\left(\frac{H_3(y)}{H_3(x)}\right)^{2/3}(dy_5^2+dy_6^2)\,.\nonumber
\eeqa

Here $ds^2_{(5)}$ is the five-dimensional metric obtained from
\reef{magnetic} after setting $N=3$, and
replacing
\beq\label{hfactor}
H(\xi)\to \left[H_1(\xi)H_2(\xi)H_3(\xi)\right]^{1/3}\,.
\eeq
The four-form field strength is
\beq
F_{[4]}=3\left(dA_{(1)}\wedge dy_1\wedge dy_2+dA_{(2)}\wedge dy_3\wedge dy_4+
dA_{(3)}\wedge dy_5\wedge dy_6\right)
\eeq
with
\beq
A_{(i)}=C(\nu,-\mu_i)\:R\:\frac{1+x}{H_i(x)}\:d\varphi\,.
\eeq
Each M5 brane spans the $\psi$ direction of the ring and four of the
$y_n$ directions.

Given that the triple intersection results in a factorized structure in
the metric coefficients, \reef{hfactor} (in spite of the absence of any preserved
supersymmetries), most results follow from simple substitutions 
in the expressions in section \ref{dipolsol}. For instance, equilibrium now requires
\beq\label{period3}
\Delta\varphi=\Delta\psi=
2\pi\frac{\sqrt{1-\lambda}}{1-\nu}\prod_{i=1}^{3}\sqrt{1+\mu_i}\,,
\eeq
and
\beq\label{equil3}
\frac{1-\lambda}{1+\lambda}\prod_{i=1}^3\frac{1+\mu_i}{1-\mu_i}=
\left(\frac{1-\nu}{1+\nu}\right)^2\,,
\eeq
and the physical parameters are
\beq\label{mass3}
M=\frac{3\pi R^2}{4G }\frac{1}{1-
\nu}\left(\lambda+\sum_{i=1}^3\frac{\mu_i(1-\lambda)}{3(1+\mu_i)}\right)
\prod_{i=1}^3(1+\mu_i)\,,
\eeq

\beq\label{spin3}
J=\frac{\pi R^3}{2G }\frac{\sqrt{\lambda(\lambda-
\nu)(1+\lambda)}}{(1-\nu)^2}\prod_{i=1}^3(1+\mu_i)^{3/2}\,,
\eeq

\beq\label{area3}
{{\cal A}_H}=8\pi^2 R^3
\frac{\sqrt{\lambda(1-\lambda^2)}}{(1-\nu)^2(1+\nu)}
\prod_{i=1}^3(1+\mu_i)(\mu_i+\nu)^{1/2}\,,
\eeq

\beq\label{temp3}
T=\frac{1}{4\pi
R}\frac{\nu^(1+\nu)}{\prod_{i=1}^3(\mu_i+\nu)^{1/2}}\sqrt{\frac{1-
\lambda}{\lambda(1+\lambda)}}\,,
\eeq

\beq\label{omega3}
\Omega=\frac{1}{R}\frac{1}{\prod_{i=1}^3(1+\mu_i)^{1/2}}\sqrt{\frac{\lambda-
\nu}{\lambda(1+\lambda)}}\,,
\eeq

\beq\label{Q3}
{\cal Q}_i=R\frac{\sqrt{\mu_i(\mu_i+\nu)(1-\lambda)}}{(1-\nu)\sqrt{1-\mu_i^2}}
\prod_{j=1}^3\sqrt{1+\mu_j}\,,
\eeq

\beq
\Phi_i=\frac{\pi R}{2G }\sqrt{\frac{\mu_i(1-\mu_i)(1-
\lambda)}{(1+\mu_i)(\mu_i+\nu)}}\prod_{j=1}^3\sqrt{1+\mu_j}\,.
\eeq
These magnitudes correspond to the
five-dimensional interpretation of the solution, hence the presence of
the five-dimensional Newton's constant $G$. If we assume that the
compact $y_n$ directions have all equal length $L$, then $G $ is
related to the eleven-dimensional coupling constant
$\kappa$ as
\beq
G =\frac{\kappa^2}{8\pi L^6}\,,
\eeq
and the number of M5 branes of each type forming the
ring is \cite{kt}
\beq\label{ni}
n_i=2\pi L^2\left(\frac{2}{\pi\kappa^2}\right)^{1/3}
{\cal Q}_i=\left(\frac{2\pi}{G}\right)^{1/3} {\cal Q}_i\,.
\eeq

Note that, in the extremal limit $\nu=0$, the mass is {\it not} the sum
of component masses (plus momentum), so the extremal ring is not a
threshold bound state. The latter only occurs in the straight string
limit, which can be obtained in a manner analogous to the previous
sections. 

\subsection{Microscopic counting of the entropy}

The straight string limit of the black ring we have just described is,
modulo dualities, the same string with three charges plus momentum used
in stringy microscopic models of four-dimensional black holes
\cite{4dbh}. The intersection of the branes can be regarded as an
``effective string", capable of supporting left and right moving
excitations, analogously to a fundamental string, but with an effective
length $n_1n_2n_3$ times longer than the circle length, so the
excitation gap is correspondingly smaller. The qualitative microscopic
arguments described at the end of section \ref{floop} therefore apply
here as well, but we can also perform a more quantitative analysis and
count the string states using statistical mechanics. We will reproduce in
this way the Bekenstein-Hawking entropy of the black ring including the
leading corrections that come from considering rings at large but finite
radius. The calculation shares many features with the study of the
extremal string-antistring system in \cite{dihole2}.

We do the analysis only for the extremal ring, and leave the discussion
of non-extremal rings for future work. We shall work at large radius,
expanding in the small parameters $\lambda$, $\mu_i$, and keeping the
first order corrections in $\lambda\sim\mu_i\sim 1/R$. The equilibrium
condition \reef{equilN} is
\beq\label{equilexp}
\lambda=\sum_i\mu_i+ O(\lambda^3)\,,
\eeq
and will be imposed in the following.

The extremal system is regarded as the ground state of the ring with
finite radius, with chiral momentum excitations at level $n_p$ on a CFT
with central charge $c=6n_1n_2n_3$, or, in the long string picture, a
$c=6$ CFT but now at level
$n_1n_2n_3n_p$. Either way the degeneracy of the microscopic
state is
\beq\label{smicro}
S_{\rm micro}=2\pi\sqrt{n_pn_1n_2n_3}\,.
\eeq
This must be compared to the Bekenstein-Hawking entropy of the ring. To
this effect, we have already identified the number of branes forming the
ring, $n_i$, in eq.~\reef{ni}. To obtain $n_p$ we want
to use a definition that captures
the momentum that runs along the ring in a manner similar to the way that
the brane numbers $n_i$ characterize the winding along the ring.
Observe that the brane
numbers $n_i$ are  
defined through an integral over a surface
that encloses a section of the ring, and the result is independent of
the specific geometry or location of the surface as
long as it encloses the ring as in fig.~\ref{fig:ringdipole}. 
A similarly covariant definition can be given for the momentum number
$n_p$ in terms of a Komar integral, analogous to those studied in
\cite{toza}. Namely, if $\eta$ is the one-form dual to the Killing
vector $\partial_{\tilde\psi}$ that generates translations along the ring,
with $\tilde\psi$ canonically normalized to have periodicity $2\pi$, \ie 
$\eta=\frac{\Delta\psi}{2\pi}g_{\psi\mu}dx^\mu$, then
we identify
\beq\label{np}
n_p=\frac{1}{16\pi G}\int_{S^1\times S^2}\ast d\eta\,,
\eeq
where the surface $S^1\times S^2$ encloses the entire ring and can be
taken to lie at constant $y\in (-1/\nu,-1)$ and constant $t$. Since
$\ast d\eta$ is regular (\ie vanishes) at the axis $y=-1$, the
integration over $S^1\times S^2$ in \reef{np} can be extended to an
integration over an $S^3$ that encloses the ring completely. This is then
the same as the Komar integral for the angular momentum contained inside
the $S^3$. For these solutions the gauge fields do not contribute to
this angular momentum, so its value is independent of the location of
the integration surface, as long as it encloses the ring, and so the
$S^3$ can be taken at asymptotic infinity. Then \reef{np} must be equal
to the ADM value of the angular momentum, \ie
\beq
n_p=J\,,
\eeq
with $J$ given in \reef{spin3}. 

This result is valid for general
non-extremal rings. If we take the extremal limit $\nu=0$ and expand for
large radius, 
\beq\label{np2}
n_p=\frac{\pi R^3}{2G}\lambda \left(1+2\lambda+O(\lambda^2)\right)\,.
\eeq
In this same limit the brane numbers \reef{ni} are
\beq\label{ni2}
n_i=\left(\frac{2\pi}{G}\right)^{1/3}R\mu_i \left(
1+O(\lambda^2)\right)\,.
\eeq
We have used \reef{equilexp} to simplify the terms in brackets in these expressions.
The Bekenstein-Hawking entropy of the extremal ring is, from
\reef{area3},
\beq
S_{BH}=\frac{{\cal A}_H}{4G}=\frac{2\pi^2
R^3}{G}\sqrt{\lambda\mu_1\mu_2\mu_3}\left(1+\lambda
+O(\lambda^2)\right)\,,
\eeq
and comparing to the microscopic entropy \reef{smicro}, using \reef{np2}
and \reef{ni2}, we find
\beq
S_{BH}=S_{\rm micro}\left( 1+O(\lambda^2)\right)\,.
\eeq
The agreement between the leading terms is of course nothing but the
result of \cite{4dbh} for a straight string, but the corrections at
first order in $\lambda$, which are also reproduced correctly, are a
genuine feature of the black ring. Like in the string-antistring system
in \cite{dihole2}, obtaining agreement beyond the leading correction
appears to require further refinements, presumably because the departure
from the supersymmetric state becomes too large.

The fact that we have reproduced the entropy using the same formula
\reef{smicro} as in the case of a straight string does not mean that
the bending of the string has no effect on the microscopic states. The
forces that are present (centrifugal and self-attraction of the ring)
produce a shift in the excitation levels, which is crucial in analyzing
the excitations above extremality. However, the
degeneracy of the ground state is not affected by these shifts.

\setcounter{equation}{0}
\section{Discussion}
\labell{discuss}

We have supplied the first example, to our knowledge, of black hole
solutions that are asymptotically flat, with regular horizons, and which
are the source of a dipolar gauge field. They also imply the violation
of uniqueness by a continuous parameter, for solutions of
\reef{emdaction} and \reef{e2daction} with 
\beq
\pi J^2> GM^3\,.
\eeq 
This is black hole hair of the
most basic type ---no ``secondary hair", nor exotic fields--- in the
familiar Einstein-Maxwell theory. It is then clear that any
notion of black hole uniqueness in the most basic theories in higher
dimensions can not be too simple. Following the discovery of the neutral
black ring, several steps have been taken towards this
goal\footnote{Ref.~\cite{kol} makes some speculations in this
direction.}. Ref.~\cite{harv} proved that the only supersymmetric black
hole of minimal five-dimensional supergravity is the BMPV solution
\cite{bmpv}. This makes it very unlikely that supersymmetric black rings with
a regular horizon exist at all in any five-dimensional supergravity theory.
Refs.~\cite{statunique} have proven the uniqueness of
static black holes. The stationary case is typically much more
complicated, but recently it has been established that the MP solution is the
unique black hole in five dimensions among the class of neutral
solutions with spherical topology and with three commuting Killing
vectors \cite{rotunique}. Since we have not found any spherical black
holes with a gauge dipole field (see the appendix), it seems likely that
this result should extend as well to the theories \reef{emdaction},
\reef{e2daction} studied in this paper. It is still unknown whether the
solutions with only two commuting Killing symmetries conjectured in
\cite{harv} actually exist, and if they do, whether they can be dipole
sources as well. Ref.~\cite{sem} began a systematic study of stationary
solutions of the Einstein-Maxwell theory in higher dimensions. Other
interesting solutions in this class have been found in \cite{teo}, and a
study of the electromagnetic properties of five-dimensional rotating
black holes has been carried out in \cite{af}.

The existence of dipole black rings in five dimensions contrasts with the
absence in four dimensions of asymptotically flat black holes, regular
on and outside the horizon, with a gauge dipole and no monopole
charge\footnote{The solutions with an electric dipole in \cite{HoTa}
have singular horizons.}. To be sure, there do exist solutions that
describe two static black holes, with charges equal in magnitude but
with opposite sign, and which therefore form a dipole
\cite{dihole1,dihole2}. But the two black holes attract each other, and
if we tried to balance the attraction by spinning the configuration,
radiation would be generated, losing stationarity. Alternatively, the
dipole can be balanced, both in four and five dimensions, by immersing
it in a Melvin-like field, \ie a fluxbrane, or with cosmic strings, but
then asymptotic flatness
is lost \cite{dihole1,tubular}.

It is interesting to compare the different string theory realizations of
black rings, in this paper and in \cite{EE}. In ref.~\cite{EE} the black
ring results from a tubular intersection of D1-branes and D5-branes,
properly viewed as a six-dimensional configuration. Reduction along the
direction of the tube $z$ results into a five-dimensional black ring
with net D1 and D5 charges. The effective string extends along $z$, \ie
transversely to the ring, and the angular momentum is provided by the
intrinsic spin of the ground state, \ie it is unrelated to the presence
of any momentum along the string. In the rings in this paper, the
effective string lies instead along the direction of the ring, and the
angular momentum is actually the result of momentum carried by
excitations moving along the circle. The extremal ground states, and the
excitations above them, are therefore
quite different in each case. Also, the rings in \cite{EE} have
supersymmetric limits with finite radius (supertubes), whereas in the
present paper supersymmetry can only be preserved in the straight string
limit. So black rings can appear in string theory from rather different
perspectives. 

Nevertheless, the microscopic picture for the neutral black ring
suggested by the analysis in \cite{EE} nicely dovetails with the idea in
this paper that a black ring can be viewed as a loop of string with
excitations running along the loop. In keeping with the string/black
hole correspondence principle \cite{corr1,corr2}, ref.~\cite{EE}
proposed that neutral black rings go over, as the string coupling is
decreased, into highly-excited fundamental strings in a fuzzy
donut-shaped configuration. Now, if we start from a dipole ring with
small winding $n$, and we add left and right moving excitations to the
string loop, we will be moving further away from extremality
(effectively decreasing $q$) and approaching a configuration more and
more similar to the fuzzy string-donut proposed for the neutral black
ring. The latter would simply correspond to the case where there is no
net winding around the loop. 

Pushing this picture further, we find an appealing explanation of the
differences between the large and small neutral ring branches
(fig.~\ref{fig:ajneu}). They would differ in the way the winding totals
to zero: rings in the small branch would contain some strings wound with
opposite orientation, whereas large rings would correspond to
configurations where no string makes a full wind. Amusingly, this simple
idea can easily explain several peculiarities of these branches of
solutions. The existence of a limit
in the small ring branch, with maximum $j$ and zero area, follows if
this limiting configuration consists of oppositely oriented pairs of
extremal strings, with only, say, left-movers. The argument is similar
to those given in section~\ref{floop}. In this configuration the total
energy $M$ has to be distributed among the tension of wound strings and
the momentum of oscillators. The presence of wound strings limits the
radius of the ring, since the winding energy grows with this radius.
Then the angular momentum, which for given amount of momentum along the
ring also grows with the radius, will also be bounded above, and the
bound saturated when the oscillators all move in the same sense -- \ie
we have pairs of oppositely wound extremal strings. The degeneracy of
extremal fundamental strings is not large enough to register in the
gravitating solution, and so this configuration appears as a zero-area
singularity. If we then add some right-movers the temperature raises, we
enhance the entropy, and a finite horizon area appears, much like when a
gravitating fundamental string becomes non-extremal. This increase in
$a_H$ should come with a decrease in $j$, due to the right-movers, and
this is precisely what we observe in fig.~\ref{fig:ajneu} as we move
along the small ring curve starting from $(j^2,a_H)=(1,0)$. 

We can also
see the reason for the smaller area of small rings: a significant
fraction of their energy has to go into the (non-entropic) tension of
wound strings, whereas for large rings the energy is mostly spent into
oscillations that contribute to the entropy. Two oppositely wound closed
strings in a small ring can intercommute and then unwind (or
annihilate). More energy is then available for oscillations, which
increase the entropy (area), and so this provides a mechanism by which
small rings decay into thermodynamically favoured large rings. Moreover,
the point at which both branches meet admits a natural interpretation:
if we approach it from the large ring branch, then we have a state of
high excitation, with unwound strings so long that they stretch almost
all the way around the ring circle. If one of these long closed strings
self-intersects after a turn around the circle, it can cut itself and
rejoin into two oppositely wound strings, thus connecting to the small
ring branch. Additionally, if we put in some strings that wind in only
one direction, we make a dipole ring. As expected by these arguments,
adding this wound strings, \ie increasing $q$, has the effect of
reducing the entropy for a given mass, again in agreement with
fig.~\ref{fig:ajn}.

This qualitative sketch provides a suggestive basis for the
identification of the kind of string states that correspond to each
different black object, \ie a complete stringy resolution of black hole
non-uniqueness in five dimensions, but there remain obscure points. One
puzzling feature is the absence of an infinite radius limit for the
configurations conjectured to correspond to small rings. More
quantitatively, it is unclear why the maximum value of $j$ for small
neutral rings is 1, and not a distribution of values as could be naively
expected if there can be configurations with different numbers of
strings and anti-strings in the ring. Or why the small ring branch for
$N=1$ terminates before reaching the extremal limit. Presumably we
require a deeper comprehension of the equilibrium of forces in the ring
to understand these issues, and decide if the picture above is tenable.
Also, the Bekenstein-Hawking entropy of boosted straight strings is
known to parametrically match, at the correspondence point, to the
entropy of highly-excited fundamental strings with linear momentum
\cite{corr2}. It might be interesting to extend this calculation to
large rings and hopefully also to small rings.

The stability of rotating dipole rings is an intriguing question, as it
is for all rotating black objects in higher dimensions
\cite{ER,EM,rotstab} (the stability problem in the static case has been
solved recently \cite{statstab}). Obviously, the dipole ring can evolve
into a neutral configuration without any gauge dipole field, since there
is no conserved quantity associated with a dipole. So, for example, the
dipole ring could collapse into a MP black hole. Another possibility is
that the dipole discharges via the formation of closed string loops
formed just outside the inner rim of the ring (presumably via quantum
effects), and which shrink to zero size radiating away their energy.
This would gradually decrease the local charge of the ring, and
depending on the amount of spin that is lost, might end up in a MP black
hole or a neutral black ring. Besides, one expects thin dipole rings to
suffer from the classical Gregory-Laflamme (GL) instability \cite{GL},
which should form ripples along the ring, and these would cause the
emission of gravitational radiation. If ${\cal Q}$ is not changed under this
process, which does not seem unlikely (no classical emission of string
loops), then $q$ would increase, and if the loss of spin were not too
large, the dipole could evolve towards an extremal solution. It is
unclear whether extremal and near-extremal rings should be GL-unstable
or not. As the supersymmetric limit is approached the GL-unstable modes
stretch to infinite wavelength. However, for these rings the
supersymmetric limit involves, in addition to extremality, taking the
radius to infinity. It remains to be seen whether extremal rings at
finite radius can be stable under classical linearized perturbations.

\section*{Acknowledgements} 
We would like to thank Henriette Elvang, Rob Myers, Harvey Reall and
Jorge Russo for conversations, comments and suggestions. Work supported
in part by grants UPV00172.310-14497, MCyT FPA2001-3598, DURSI
2001-SGR-00188, and HPRN-CT-2000-00131. 

\appendix

\setcounter{equation}{0}
\section{Limit of Myers-Perry black hole}
\label{app:bhlimit}

In order to recover the five-dimensional MP black hole with rotation in one
plane \cite{MP} from the solution \reef{neutral}, define new parameters
$a$, $m$,
\beq
%\lambda=1-2\left(1-\frac{a^2}{m}\right)\frac{R^2}{m}\,,
%\qquad \nu=1-2\:\frac{R^2}{m}
m=\frac{2R^2}{1-\nu}\,,\qquad a^2=2R^2\frac{\lambda-\nu}{(1-\nu)^2}\,,
\eeq
such that they remain finite as $\lambda,\nu\to 1$ and $R\to 0$. Also,
change coordinates
$(x,y)\to (r,\theta)$,
\beqa
x&=&-1+2\left(1-\frac{a^2}{m}\right)\frac{R^2\cos^2\theta}{r^2-(m-
a^2)\cos^2\theta}\,,\nonumber\\
y&=&-1-2\left(1-\frac{a^2}{m}\right)\frac{R^2\sin^2\theta}{r^2-(m-
a^2)\cos^2\theta}\,,
\eeqa
and rescale $\psi$ and $\varphi$
\beq
(\psi,\varphi)\to \sqrt{\frac{m-a^2}{2R^2}}\;(\psi,\varphi)
\eeq
so they now have canonical periodicity $2\pi$. Then we recover the metric
\beq
ds^2=-\left(1-\frac{m}{\Sigma}\right)\left(dt-\frac{m
a\sin^2\theta}{\Sigma-
m}\:d\psi\right)^2+\Sigma\left(\frac{dr^2}{\Delta}+d\theta^2\right)
+\frac{\Delta\sin^2\theta}{1-
m/\Sigma}\:d\psi^2+r^2\cos^2\theta\:d\varphi^2\,,
\eeq
\beq
\Delta\equiv r^2-m+a^2\,,\qquad \Sigma\equiv
r^2+a^2\cos^2\theta
\eeq
of the MP black hole rotating in the $\psi$ direction.

Consider now this same limit for the dipole solution \reef{magnetic}.
Focusing on the gauge field and taking $\lambda,\nu\to 1$ and
$R\to 0$, there does not appear to be any way to obtain a limiting
solution with finite horizon and a non-trivial gauge field. So we
can only recover the MP solution, and we do not find any new black hole
of spherical topology with dipole charge.

%%%%%%%%%%%%%%%%%%%%%%%%%%%%%%%%%%%%%%%%%%%%%%%%%%%%%%%

\end{document}